\documentclass[aip,jcp,reprint,superscriptaddress,letterpaper,longbibliography]{revtex4-2}

\usepackage{amsmath}
\usepackage{amssymb}
\usepackage{pifont} 
\usepackage{braket}
\usepackage{mathtools}
\usepackage{bm}
\usepackage{siunitx}
\usepackage {verbatim}

\usepackage{hyperref}
\hypersetup{
    colorlinks=true,
    linkcolor=black, 
    citecolor=black, 
    filecolor=black, 
    urlcolor=black 
}
\usepackage[table]{xcolor} 
\usepackage{colortbl}

\begin{document}
\title {Determining the complex second-order optical susceptibility in macroscale van der Waals heterobilayers}
\author{Zeyuan Zhu}
\thanks{These authors contributed equally to this work.}
\affiliation 
{Department of Chemistry, University of Michigan, Ann Arbor, MI 48109, USA}
\author{Taejun Yoo}
\thanks{These authors contributed equally to this work.}
\affiliation 
{Department of Chemistry, University of Michigan, Ann Arbor, MI 48109, USA}
\author{Kanchan Shaikh}
\thanks{These authors contributed equally to this work.}
\affiliation 
{Department of Chemistry, University of Michigan, Ann Arbor, MI 48109, USA}
\author{Amalya C. Johnson}
\affiliation 
{Department of Materials Science and Engineering, Stanford University, Stanford, CA 94305, USA}
\author{Qiuyang Li}
\affiliation{Department of Physics, University of Michigan, Ann Arbor, MI 48109, USA}
\author{Fang Liu}
\affiliation {Department of Chemistry, Stanford University, Stanford, CA 94305, USA}
\author{Hui Deng}
\affiliation{Department of Physics, University of Michigan, Ann Arbor, MI 48109, USA}
\affiliation{Department of Electrical Engineering and Computer Science, University of Michigan, Ann Arbor, MI 48109, USA}
\author{Yuki Kobayashi}
\email{ykb@umich.edu}
\affiliation {Department of Chemistry, University of Michigan, Ann Arbor, MI 48109, USA}
\date{\today}
\begin{abstract}
We report on the experimental characterization of the second-order susceptibility in MoSe$_2$/WS$_2$ heterobilayers, including their hidden complex phases.
To this end, we developed a heterodyne-detection scheme for second-harmonic generation and applied it to macroscale heterobilayer samples prepared using the gold-tape exfoliation method.
The heterodyne scheme enabled us to distinguish the relative orientation of the crystal domains, and further, it allowed us to characterize the complex phases of the susceptibility relative to a reference quartz sample.
By comparing the results from the monolayer regions and the heterobilayer region over several hundred microns of the sample area, we determined that the contribution of interlayer effects to second-harmonic generation is within the experimental uncertainty arising from the sample inhomogeneity.
The results here provide fundamental quantitative information necessary for the precise design of nanophotonic systems based on stacking engineering.
\end{abstract}
\maketitle
\section{Introduction}
Two-dimensional materials, including transition-metal dichalcogenides (TMDs), exhibit enhanced per-layer optical properties due to their reduced dimensionality. \cite{manzeli_2d_2017,tebyetekerwa_mechanisms_2020,mennel_second_2018,wang_nonlinear_2015}
Furthermore, stacking engineering of two-dimensional materials, achieved with atomic precision, can offer unique opportunities to control their optical and electrical properties. \cite{2016m,ma_harmonic_2021,zhou_controlling_2022,jin_ultrafast_2018,lu_twisted_2017,sebait_sequential_2023}
As a result, two-dimensional materials represent a promising platform for nonlinear nanophotonics,\cite{pal_quantum-engineered_2023,yu_2d_2017} enabling applications in high-harmonic generation,\cite{heide_high-harmonic_2023,liu_high-harmonic_2017,yoshikawa_interband_2019} optical parametric amplification,\cite{trovatello_optical_2021} and high-order sideband generation.\cite{langer_lightwave_2018}

Second-harmonic generation (SHG) in monolayer TMDs has been extensively studied, with research primarily focusing on determining the magnitude of the second-order susceptibility.\cite{malard_observation_2013,li_probing_2013,kumar_second_2013}
The second-harmonic signals are sensitive to the crystal symmetry and electronic structures and have been used as a powerful spectroscopic tool to probe a wide range of phenomena, including ferroelectricity,\cite{denev_probing_2011} ferro-rotational order, \cite{jin_observation_2020} interlayer coupling, \cite{yuan_probing_2023,le_effects_2020} and interlayer excitons. \cite{shree_interlayer_2021,kim_exciton-sensitized_2023}
However, studies investigating the complex phase of the second-order susceptibility are limited.
Complex-phase information of optical properties becomes essential when interference from multiple contributions needs to be considered.\cite{ohno_phase-referenced_2016, aghigh_second_2023}
Second harmonic generation from multi-layer crystals of two-dimensional materials is a prime example of this case.
Previous research on artificially stacked TMDs homostructures showed a coherent enhancement of the second-harmonic signals with increasing layers up to five, attributed to negligible phase mismatch.\cite{liu_disassembling_2020} 
In contrast, when the stacked layers are composed of different materials, it is necessary to explicitly consider the complex phases in the second-order susceptibility, a topic that remains unexplored. 

To address this knowledge gap, we employ the second-harmonic heterodyne detection to resolve the relative phase information and stacking orientations in the macroscale MoSe$_2$/WS$_2$ heterobilayer.
Using this technique, we further characterize the magnitude and complex phases of the second-order susceptibility tensor in both monolayers and heterobilayer.
These comprehensive measurements enable a direct comparison between the independent-layer model and the experimental results for heterobilayer, determining the upper limit of the potential contribution from the interlayer interactions.
The results show that the effects of the interlayer interactions are negligibly small on the order of less than 1\%, whereas sample inhomogeneity, particularly in signal magnitude but not phase, may contribute a few percent of signal deviation.
\section{Results and Discussion}
\label{Results and Discussion}
\subsection{Sample preparation and experimental setup}
\begin{figure*}
 \includegraphics{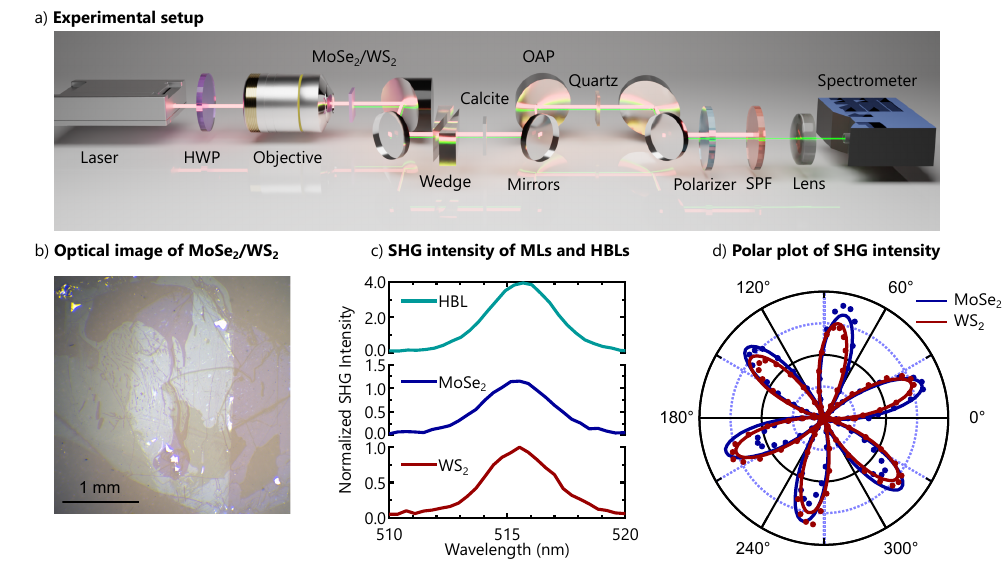}
 \caption{
\textbf{An overview of the heterodyne second-harmonic generation experiments.}
 (\textbf{a}) A schematic illustration of the optical layout.
 We measure interference between the second-harmonic generation from TMDs and z-cut quartz, with phase precisely controlled by a fused silica wedge pair.
 The TMD sample is mounted on a two-axis motorized stage for precise control.
 (\textbf{b}) An optical image of the macroscale sample with monolayer, in which MoSe$_2$ is located on the left, monolayer WS$_2$ on the right, and heterobilayers MoSe$_2$/WS$_2$ at the center.
 (\textbf{c}) Second-harmonic spectra from WS$_2$, MoSe$_2$, and MoSe$_2$/WS$_2$, normalized by the monolayer WS$_2$ second-harmonic signal. 
 (\textbf{d}) Polarization-dependent second-harmonic measurement of monolayers MoSe$_2$ and WS$_2$.
 The result shows that the stacking angle of the two monolayers is 2.6$^{\circ}$.}
 \label{fig:Fig1}
\end{figure*}
The heterobilayer samples were fabricated by using the gold-tape exfoliation method, \cite{liu_disassembling_2020} and a summary is as follows.
We first prepared self-assembled monolayers of 1-dodecanol as a passivation layer to the fused-silica substrates, thus enhancing the surface flatness and optical properties of the TMD samples.\cite{li_macroscopic_2023} 
We then prepared the monolayer WS$_2$ on the passivated substrate using mechanical exfoliation with gold tapes.
The exfoliated monolayer MoSe$_2$ was subsequently stacked onto the monolayer WS$_2$ by aligning the crystal edges under a microscope.
Compared to the traditional scotch-tape exfoliation method, the gold-tape exfoliation method allows us to prepare millimeter-scale monolayer TMDs, making them well-suited for flexible integration into laser spectroscopy setups.
Figure \ref{fig:Fig1}\textbf{b} shows an optical image of the macroscopic millimeter-scale MoSe$_2$/WS$_2$ heterobilayer. 
We determined the stacking angle of 2.6$^\circ$ for the MoSe$_2$/WS$_2$ sample by performing the polarization-dependent second-harmonic measurements on the individual monolayers, as shown in Fig. \ref{fig:Fig1}\textbf{d}. 

The homodyne and heterodyne detections of the second-harmonic signals were performed using a home-built beamline equipped with a femtosecond laser source (1030 nm, 200 fs, 100 mW, 40 MHz), as depicted in Fig. \ref{fig:Fig1}\textbf{a}. 
For the homodyne detection, we focused the 1030 nm laser on the TMDs by a microscope objective (10$\times $, NA 0.25, Plan) with a spatial resolution of 18 \SI{}{\micro\metre}. 
Figure \ref{fig:Fig1}\textbf{c} presents the second-harmonic spectra from monolayer MoSe$_2$, monolayer WS$_2$, and heterobilayer MoSe$_2$/WS$_2$,
all normalized to the second-harmonic signal intensity of monolayer WS$_2$.
The intensities of the monolayer MoSe$_2$ and the heterobilayer regions are found to be 1.1 and 3.9 times greater than that of monolayer WS$_2$, respectively.
 
To measure the second-order optical susceptibility $\chi^{(2)}$ using heterodyne detection, we generated a reference second-harmonic signal by further focusing the pump laser onto a 100-\SI{}{\micro\metre}-thick z-cut quartz plate using an off-axis parabolic mirror.
A 500-\SI{}{\micro\metre} calcite plate was placed before the reference quartz to adjust the temporal overlap of the fundamental and the second-harmonic signals.
A pair of fused-silica wedges was employed as a phase-shifting unit to precisely control the relative phase between the second-harmonic signals from the TMD samples and the reference quartz plate.
To calibrate the TMD-quartz heterodyne signals, we replace the TMD samples with a second z-cut quartz plate with 100-\SI{}{\micro\metre} thickness and perform the heterodyne detection with two quartz plates. 

Here, we express the second-order optical susceptibility as $\chi^{(2)}$=$\mid$$d$$\mid$\(\cdot\)$e^{i\phi}$, including both the magnitude of the second-order optical susceptibility tensor $d$ and the complex phase $\phi$. 
For TMD monolayers and heterobilayers with a stacking angle close to 0$^\circ$, $d$ corresponds specifically to the $d_{22}$ tensor component.\cite{jung_simple_2000} 
By comparing the heterodyne signals between the TMD-quartz and quartz-quartz configurations, we can characterize both the second-order susceptibility tensor and the complex phase for the monolayers and heterobilayers. 
The details of the analysis methods are provided in Supplemental Section I.

\subsection{Crystallographic orientations in monolayers}
\begin{figure}[htbp]
 \centering
 \includegraphics[scale = 1]{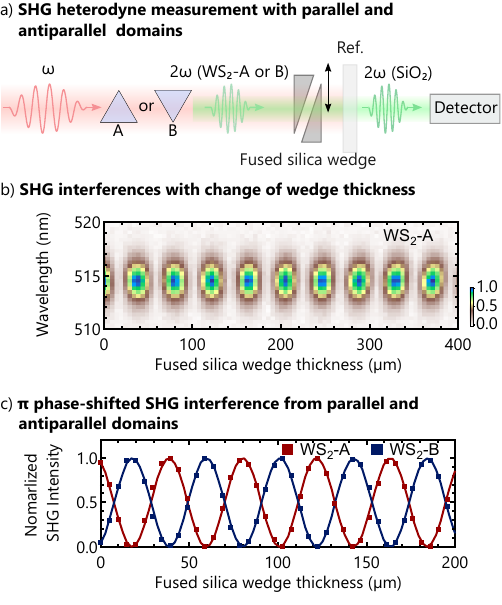}
 \caption{
 \textbf{Heterodyne detection of SHG signals from WS$_2$ monolayer with opposite crystal orientations.}
 (\textbf{a}) A schematic illustration of the second-harmonic heterodyne detection between the monolayer WS$_2$ with parallel or antiparallel subdomains and the quartz.
 (\textbf{b}) The signal interference between the monolayer WS$_2$ and the reference quartz as a function of fused silica wedge thickness. 
 (\textbf{c}) The interference signal intensities with two different subdomains ($A$ and $B$) of WS$_2$.
 The results reveal a relative phase shift of $\pi$ between the two antiparallel domains.
 }
 \label{fig:Fig2}
\end{figure}

As an initial test of the heterodyne detection, we used it to resolve the crystal orientations within the monolayer regions (Fig. \ref{fig:Fig2}\textbf{a}).
In monolayer TMDs, the crystal surface can display parallel and antiparallel subdomains, typically represented as equilateral triangle domains (denoted as $A$) and inverted equilateral triangle domains (denoted as $B$). 
The parallel and antiparallel domains are indistinguishable in homodyne measurements as they exhibit the same second-harmonic intensities.
On the other hand, heterodyne detection can measure the complex phases of the second-harmonic signals, where the subdomain information is encoded, thereby enabling experimental separation of the crystal subdomains.\cite{wang_contrast-enhanced_2024}

Figure \ref{fig:Fig2}\textbf{b} shows the modulation of the second-harmonic signals from the domain $A$ as the thickness of the fused-silica wedge pairs is varied.
A cosine fit of the integrated signals from the parallel and antiparallel domains is shown in Fig. \ref{fig:Fig2}\textbf{c}, revealing a relative phase shift of $\pi$ between the two domains.
The presence of parallel and antiparallel domains in mechanically exfoliated TMDs can be attributed to alternating crystal orientations in consecutive layers,\cite{saunders_direct_2024} which may result from repeated exfoliation steps.  
Visualization of parallel and antiparallel domains in monolayers WS$_2$ and MoSe$_2$ is shown in the heterodyne mappings in Fig. \textbf {S1}.
\subsection{Crystallographic orientations in heterobilayers}
\begin{figure*}[t]
 \includegraphics{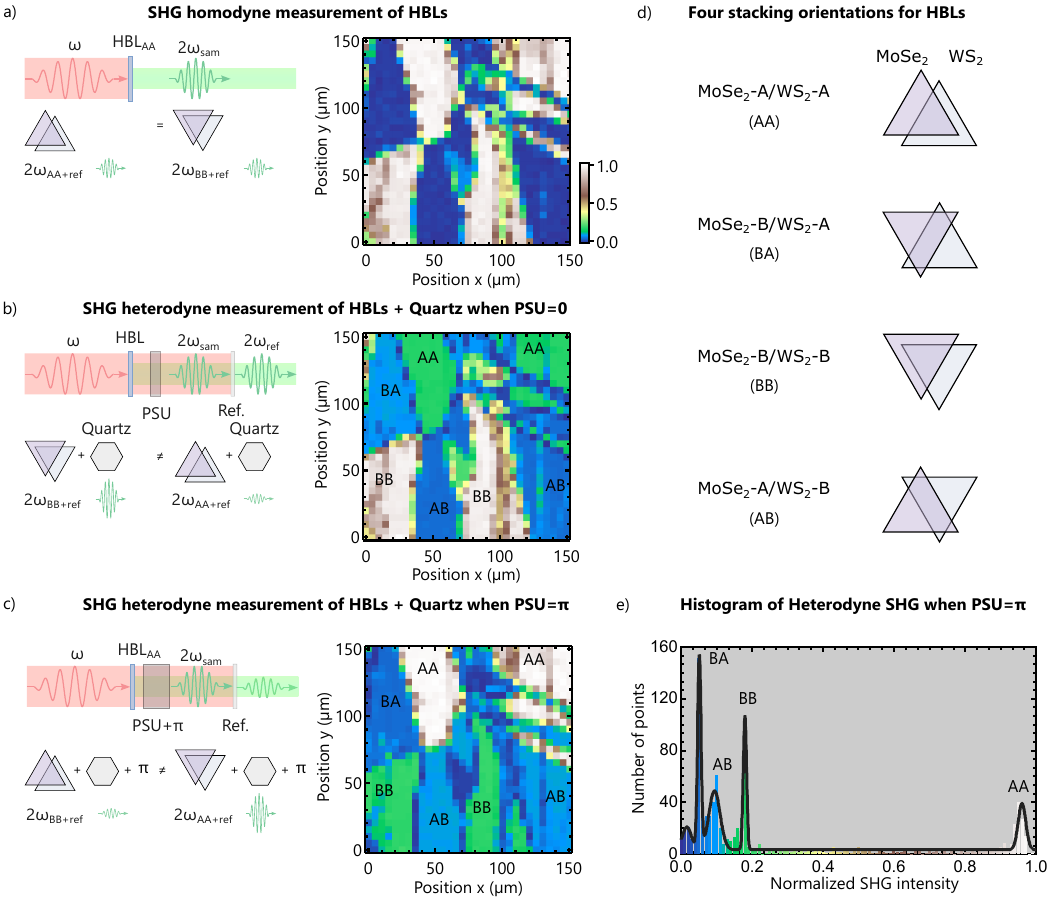}%
 \caption{
 \textbf{Heterodyne second-harmonic characterization of TMD heterobilayers.}
 (\textbf{a}) The result of the homodyne second-harmonic detection.
 This approach is insensitive to the relative phase of the second-harmonic signals.
 (\textbf{b}) The result of the heterodyne second-harmonic detection.
 Here, relative crystal-phase information is clearly resolved.
 (\textbf{c}) The result of the heterodyne second-harmonic detection with an additional $\pi$ phase shift. 
By comparing the results between (b) and (c), we can unambiguously identify the parallel ($AA$ and $BB$) and antiparallel ($AB$ and $BA$) regions.
(\textbf{d}) The four stacking orientations of heterobilayers MoSe$_2$/WS$_2$: $AA$, $BB$, $AB$, and $BA$.
(\textbf{e}) Histogram of the second-harmonic signal intensities in heterodyne detection in (c) when $\text{PSU}=\pi$, revealing four distinct peaks corresponding to the stacking orientations in (d).}
  \label{fig:Fig3}
\end{figure*}

We then move on to the stacked regions of heterobilayer MoSe$_2$/WS$_2$.
Corresponding to the $A$ and $B$ subdomains in the monolayers, the macroscale heterobilayers consist of four distinct stacking orientations, as shown in Fig. \ref{fig:Fig3}\textbf{d} and labeled as $AA$, $AB$, $BA$, and $BB$.
These configurations differ in their relative crystallographic orientations, resulting in constructive or destructive interference of the second-harmonic signals depending on the phase alignment between the two layers.
Specifically, $AA$ and $BB$ correspond to constructive interference resulting from in-phase stacking, while $AB$ and $BA$ correspond to destructive interference arising from out-of-phase stacking.

Figure \ref{fig:Fig3}\textbf{a} shows the second-harmonic homodyne mapping of the heterobilayer, in which the phase information of the second-harmonic signal is absent.
In this data, the two constructive configurations ($AA$ and $BB$) and the two destructive configurations ($AB$ and $BA$) are indistinguishable.

In contrast, the heterodyne mapping shown in Figs. \ref{fig:Fig3}\textbf{b} and \ref{fig:Fig3}\textbf{c} clearly resolve the four different domains.
A controlled $\pi$ phase shift is introduced from Fig. \ref{fig:Fig3}\textbf{b} to Fig. \ref{fig:Fig3}\textbf{c} by increasing the thickness of the fused-silica wedge pairs.
Under these conditions, the heterodyne signals from the $AA$/$AB$ domains with a phase shift of $\pi$ become equal to the signals from the flipped $BB$/$BA$ domains, allowing for clear identification of the stacking orientations. 
Figure \ref{fig:Fig3}\textbf{e} presents a histogram analysis of the heterodyne intensity in Fig. \ref{fig:Fig3}\textbf{c}, where the four distinct peaks correspond to the four stacking orientations in the heterobilayers.
\subsection{Characterization of second-order susceptibility $d_{22}$ and complex phase $\phi$ in monolayers and heterobilayers}
With the crystal subdomains identified in the monolayers and heterobilayers, we are now ready to extract the second-order susceptibility tensor along with its complex phase.
Our main focus here is determining the upper limit of the interlayer interactions in heterobilayers that could potentially contribute to second-harmonic generation.
To this end, by using the obtained information on the complex phases, we estimate the second-order susceptibility of the heterobilayers by taking a coherent superposition of the individual monolayers, assuming no interlayer effects (Fig. \textbf{S3}).
We then use the difference between the individual-monolayer model and the actual heterobilayer results as a measure of the interlayer effects.
We note that the individual-monolayer model includes a correction for approximately 6$\%$ absorption of the second-harmonic signal from the top MoSe$_2$ layer by the bottom WS$_2$ layer (Supplemental Section I). \cite{li_measurement_2014}
\begin{table}[]
\begin{tabular}{|ccc|ccc|}
\hline
\multicolumn{3}{|c|}{Individual monolayers} & \multicolumn{3}{c|}{Heterobilayers}\\
\hline
\multicolumn{1}{|c|}{}  & \multicolumn{1}{c|}
{\begin{tabular}[c]{@{}c@{}}$\phi'$\\ (rad.)\end{tabular}} & \begin{tabular}[c]{@{}c@{}}$d_{22}'$ $\times$ 10$^{-21}$\\
(m$^2$/V)\end{tabular} & \multicolumn{1}{c|}{}     & \multicolumn{1}{c|}{\begin{tabular}[c]{@{}c@{}}$\phi$\\ (rad.)\end{tabular}} & \begin{tabular}[c]{@{}c@{}}$d_{22}$ $\times$ 10$^{-21}$\\
(m$^2$/V)\end{tabular} \\
\hline
\multicolumn{1}{|c|}{M$_A$W$_A$} & \multicolumn{1}{c|}{\begin{tabular}[c]{@{}c@{}}-0.45\\
(0.02)\end{tabular}} & {\begin{tabular}[c]{@{}c@{}}9.10\\
(0.15)\end{tabular}} & \multicolumn{1}{c|}{M$_A$W$_A$} & \multicolumn{1}{c|}{\begin{tabular}[c]{@{}c@{}}-0.48\\
(0.01)\end{tabular}} & {\begin{tabular}[c]{@{}c@{}}9.05\\
(0.03)\end{tabular}}                        \\
\cline{1-2} \cline{4-5} \multicolumn{1}{|c|}{M$_B$W$_B$} & \multicolumn{1}{c|}{\begin{tabular}[c]{@{}c@{}}2.63\\
(0.02)\end{tabular}}  &   & \multicolumn{1}{c|}{M$_B$W$_B$} & \multicolumn{1}{c|}{\begin{tabular}[c]{@{}c@{}}2.60\\
(0.01)\end{tabular}}  & \\
\hline
\end{tabular}
\caption{
\textbf{Experimentally determined values of the second-order susceptibilities.}
Summary of second-order susceptibility tensor along with the complex phase for both predicted and experimental heterobilayers.
The values in parentheses represent the experimental standard deviations.
}
\label{table1}
\end{table}

Table \ref{table1} summarizes the experimentally measured and predicted values of the second-order susceptibility tensor and the complex phase for the heterobilayer MoSe$_2$/WS$_2$.
Our analysis is limited to the constructive stacking configurations of $AA$ and $BB$.
The destructive stacking configurations of $AB$ or $BA$ are not discussed in the following sections due to the low second-harmonic counts.

The results indicate that the difference between the individual-monolayer model and the experimental heterobilayer results is within the uncertainty of the measurements, on the order of less than 1\%.
The largest source of uncertainty arises from sample inhomogeneity, an important factor of consideration when scaling the crystal sizes to the millimeter range.
Specifically, we found that the complex phases are robust over several hundred microns of the sample area, showing an uncertainty of 40 mrad.
This deviation translates to less than 0.1\% variation when plugged into $\cos(\phi)$, virtually making no difference in our calculations.
On the other hand, the uncertainty in the absolute magnitude can be non-negligible, reaching up to 2\% of uncertainty in our samples.
Overall, our comprehensive experimental analysis shows that the potential contribution from the interlayer effect has an upper limit of 1\% for our experimental conditions.

Before closing, we would like to make a few additional remarks on interlayer effects.
First, interlayer interactions are generally known to be significant in MoSe$_2$/WS$_2$ heterobilayers with stacking angles below 5$^\circ$.\cite{alexeev_resonantly_2019}
In addition, we performed photoluminescence measurements of the heterobilayer WS$_2$/MoSe$_2$ and monolayer MoSe$_2$ at 5 K (Fig. \textbf{S2}).
The results support the possible quasi-type II band alignment of our heterobilayer samples, leading to the formation of interlayer hybrid exciton and trion bands.\cite{lu_twisted_2017,vvu_exploring_2024}
\section{Conclusions}
\label{Conclusions}
In this work, we performed heterodyne detection of second-harmonic generation and resolved the relative phases of the TMD monolayers and heterobilayers. 
The heterodyne technique enabled us to map out the parallel and antiparallel crystallographic subdomains in the monolayers and further identified the four possible stacking orientations in the heterobilayers.
Additionally, we measured the second-order susceptibility tensors, including their complex phases for both monolayers and heterobilayers, key information missing from conventional homodyne-detection experiments.
With the complex second-order susceptibility characterized, we compared the results between the individual-layer model and the actual measurements from the heterobilayers.
We determined that the potential contribution of interlayer effects to second-harmonic generation in heterobilayer samples is smaller than the impact of sample inhomogeneity, accounting for less than 1\%.
This provides an important guideline for stacking engineering of macroscale two-dimensional photonics: while interlayer interactions might be insignificant to alter their performance, sample inhomogeneity should be considered.
Meanwhile, we found that complex phase information is robust across sample sizes of several hundred microns, making it reliable for determining crystal orientation.
Lastly, it bears mentioning that our measurements were conducted at the driving wavelength of 1030 nm.
Different behaviors may show up at different wavelengths, especially when they are resonant with interlayer excitonic features.
\section*{Methods}
\label{Methods}
\subsection{Substrate passivation with 1-dodecanol self-assembled monolayers}
We prepared the 1-dodecanol passivation layer following a procedure similar to that previously reported.\cite{li_macroscopic_2023}
The 1-dodecanol was drop-casting at 160$^\circ$ on $\sim$10 mm × 10 mm × 0.5 mm fused silica substrates for 2 minutes. 
Then, the extra 1-dodecanol was washed with isopropanol and blown dry with a nitrogen gun. 
The self-assembled monolayer was chemically adsorbed onto the substrate. 
This increases the surface's flatness and can largely suppress the substrate-induced non-radiative decay, which enhances the photoluminescence intensity and interlayer interactions.

\subsection{Gold-tape exfoliation of transition-metal dichalcogenide monolayers and heterobilayers}
We used the gold-tape exfoliation method\cite{liu_disassembling_2020,li_macroscopic_2023} to prepare the millimeter-scale MoSe$_2$/WS$_2$ heterobilayers on the passivated fused-silica substrate.
A 150-nm-thick gold layer was deposited on silicon substrates using electron-beam evaporation at a deposition rate of 0.05 nm/s.
A 10$\%$ wt poly-vinylpyrrolidone solution was prepared in a 1:1 mixture of ethanol and acetonitrile to protect the gold film using the spin-coating technique at 1500 rpm for 2 minutes, with an acceleration of 500 rpm/s.
After spin coating, the heat-release tape was applied to the poly-vinylpyrrolidone-protected gold film and used to peel the gold layer off the silicon substrate. 
To prepare WS$_2$ monolayers, the detached gold film was then quickly brought into contact with the surface of bulk WS$_2$, enabling exfoliation of macroscopic, millimeter-scale monolayers.
After transferring the exfoliated monolayer onto a fused-silica substrate, the sample was placed on a hot plate at 135$^\circ$C to remove the heat-release tape. 
The poly-vinylpyrrolidone layer was then removed by soaking the sample in deionized water for 4 hours, followed by soaking in an acetone bath for 1 hour.
The gold film was etched using an iodine/potassium iodide solution prepared by dissolving 2.5 g of iodine and 10 g of potassium iodide in 100 mL of deionized water.
Finally, the WS$_2$ monolayer on the fused-silica substrate was rinsed with deionized water, soaked in an isopropanol bath for 30 minutes, and dried using a nitrogen-gas stream.
We then used the same protocol to exfoliate MoSe$_2$ monolayer, and transferred the MoSe$_2$ onto the WS$_2$ monolayers under an optical microscope.
The crystal edges of the prepared WS$_2$ on the substrate and the MoSe$_2$ monolayer on gold were matched to give a stacking angle within $5^\circ$.
The second-harmonic polarization-dependent measurements further confirmed the stacking angle with an uncertainty $\pm1^\circ$.
\begin{acknowledgements}
Z.Z., T.Y., K.S., and Y.K. acknowledge the support from startup funds by the University of Michigan.
Q.L. and H.D. acknowledge the support by the Army Research Office (W911NF-25-1-0055) and the Betty and Gordon Moore Foundation (GBMF10694).  
A.C.J. acknowledges support from the Stanford Tomkat fellowship.
The preparation of part of the samples is supported by the Defense Advanced Research Projects Agency (DARPA) under Agreement No. HR00112390108.
The preparation of the gold thin films was performed at the University of Michigan Lurie Nanofabrication Facility (UM-LNF). 
\end{acknowledgements}
\bibliographystyle{apsrev4-2.bst} 
\bibliography{Bibliography}
\end{document}